\documentclass[prd,onecolumn,superscriptaddress,aps,nofootinbib,11pt]{revtex4}

\pdfoutput=1
\usepackage[utf8]{inputenc}
\usepackage{amsmath,amssymb,mathrsfs}
\usepackage{dsfont}
\usepackage{graphicx}
\usepackage{enumerate}
\usepackage{multirow}
\usepackage[usenames,dvipsnames]{xcolor} 
\usepackage{hyperref}
\hypersetup{colorlinks=true,
		breaklinks=true,
    linkcolor=Blue,          % color of internal links
    citecolor=Plum,        % color of links to bibliography
    urlcolor=YellowOrange           % color of external links
}

\usepackage[]{units}
\usepackage{ulem} 

\usepackage{bbm}
\providecommand{\id}{{\mathbbm{1}}} %scalable

\providecommand{\hY}{\hat{Y}}

\providecommand{\ZZ}{\mathbb{Z}}

\providecommand{\mss}[1]{\mbox{\scriptsize $#1$}}

\providecommand{\tp}{{\mss{\mathsf{T}}}}

\providecommand{\to}{\rightarrow}
\providecommand{\tY}{\tilde{Y}}

\DeclareMathOperator{\diag}{diag}

\usepackage{tabu}

\providecommand{\eq}[1]{\begin{equation} #1 \end{equation}}
\providecommand{\eqali}[1]{\begin{equation}\begin{aligned} #1
    \end{aligned}\end{equation}}

\providecommand{\lag}{\mathcal{L}}

\providecommand{\ZZ}{\mathbb{Z}}

\providecommand{\mss}[1]{\mbox{\scriptsize $#1$}}

\providecommand{\tp}{{\mss{\mathsf{T}}}}

\providecommand{\cM}{\mathcal{M}}

\DeclareMathOperator{\re}{\mathrm{Re}} 
\DeclareMathOperator{\im}{\mathrm{Im}} 

\providecommand{\cY}{\mathcal{Y}}

\providecommand{\eps}{\epsilon}

\providecommand{\to}{\rightarrow}

\providecommand{\btheta}{\bar{\theta}}
\begin{document}
%%%%%%%%%%%%%%%%%%%%%%%%%%%%%%%%%%%%%%%%%%%%%%%%%
\title{Seesaw Limit of the Nelson-Barr Mechanism}

\author{A.~L.~Cherchiglia}%
\email{alche@unicamp.br}
\affiliation{Instituto de Física Gleb Wataghin, \\Universidade Estadual de Campinas, Campinas-SP, Brazil}

\author{A.~G.~Dias\footnote{This paper is dedicated to the memory of our friend and collaborator,
Prof. Alex Gomes Dias (1977–2024). He will be fondly remembered for his remarkable knowledge, tireless dedication to science and generous
heart.}}%
\email{alex.dias@ufabc.edu.br}
\affiliation{Centro de Ci\^encias Naturais e Humanas, Universidade Federal do ABC, Santo Andr\'e-SP, Brasil}

\author{J.~Leite}%
\email{julio.leite@ific.uv.es} 
\affiliation{ AHEP Group, Institut de F\'{i}sica Corpuscular --
  CSIC/Universitat de Val\`{e}ncia, Parc Cient\'ific de Paterna.\\
 C/ Catedr\'atico Jos\'e Beltr\'an, 2 E-46980 Paterna (Valencia) - Spain}

\author{C.~C.~Nishi}%
\email{celso.nishi@ufabc.edu.br}
\affiliation{Centro de Matem\'atica, Computa\c{c}\~ao e Cogni\c{c}\~ao, Universidade Federal do ABC,  Santo Andr\'e-SP, Brasil}

\date{\today}

%%%%%%%%%%% ABSTRACT %%%%%%%%%%%%%%%%%%%%%%%%%%%%%%%%
\begin{abstract}
We investigate how the solution to the strong CP problem and the explanation for the observed fermion mass hierarchies can be intrinsically related. Specifically, we explore the Nelson-Barr mechanism and identify its ``seesaw limit'', where light quark masses are suppressed by large CP-violating terms. Upon adding three (two) vector-like quarks that mix with the down-type (up-type) quark sector of the Standard Model, we demonstrate how the lack of CP violation in the strong sector and the observed quark mass hierarchy can be simultaneously achieved.
\end{abstract}

%%%%%%%%%%%%%%%%%%%%%%%%%%%%%%%%%%%%%%%%%%%%%%%%%
\maketitle
%%%%%%%%%%%%%%%%%%%%%%%%%%%%%%%%%%%%%%%%%%%%%%%%%
\newpage

\section{Introduction}\label{sec:intro}

The simplicity of the gauge sector of the Standard Model (SM) ---only three parameters govern the interactions among its many fermions and gauge bosons--- stems from 
the simplicity of the SM gauge group $\mathcal{G}_{SM} = SU(3)_C \otimes SU(2)_L \otimes U(1)_Y$ and, more generally, from the structure of gauge theories.
In contrast, explaining the SM flavour sector requires at least 20 parameters once neutrino masses are included, and no underlying guiding principle is known. Our lack of understanding of the non-trivial hierarchies observed among the flavour parameters constitutes the so-called flavour puzzle\,\cite{Feruglio:2015jfa}.

Another well-known issue for which the SM does not offer a compelling explanation is the strong CP problem, i.e., the non-observation of CP violation in the strong interacting sector to a very high degree in spite of the possible presence of the effective phase $\bar{\theta}$ breaking the symmetry.
This becomes even more intriguing since $\bar{\theta}$ receives a contribution from the flavour sector, more precisely from the quark Yukawa matrices, where CP is known to be violated.

The flavour puzzle may be addressed in many ways.
For instance, the Froggatt-Nielsen mechanism\,\cite{Froggatt:1978nt} is a popular approach which explains hierarchical fermion masses from the distinctive interactions between fermions and flavons controlled by attributing different abelian flavour-dependent charges.
Another approach, more successful for leptons, is to use properly broken flavour symmetries to explain the peculiar leptonic mixing pattern; see, e.g., Refs.\,\cite{Altarelli:2010gt,Feruglio:2019ybq} for a review.
In particular, modular flavour symmetries\,\cite{Feruglio:2017spp} lead to very restrictive models.

Yet another appealing alternative is the implementation of seesaw scenarios, originally proposed to explain the smallness of neutrino masses \cite{Minkowski:1977sc, Yanagida:1979as, Gell-Mann:1979vob, Mohapatra:1979ia, Schechter:1980gr}, and later extended to charged lepton and quark masses \cite{Berezhiani:1985in,Chang:1986bp, Rajpoot:1986nv,Davidson:1987mh,Davidson:1987tr}. 
Seesaw mechanisms for quarks generally occur in models where (i) new heavy quarks that mix with the SM quarks are introduced, and (ii) the conventional Yukawa interactions between the SM quarks and Higgs are forbidden by symmetries, {\it e.g.}, in left-right \cite{Chang:1986bp, Rajpoot:1986nv,Davidson:1987mh,Davidson:1987tr}, 3-3-1 \cite{Dias:2020ryz, Dias:2022hbu} and $U(1)$ \cite{Jana:2021tlx,Mohanta:2022seo,Chen:2022wvk} extensions of the SM. 
The large masses of the extra quarks are responsible for suppressing the SM quark masses with respect to the electroweak scale as well as providing small mixing between the SM and the heavy quarks.
The presence of extra quarks that mix with the SM fields, such as the seesaw-mediating ones, leads to rich phenomenology. If sufficiently light, they can be abundantly produced and subsequently decay at colliders due to their interactions with SM gauge bosons. Additionally, their mixing with the SM quarks induces flavour-violating interactions as well as departures from unitarity in the quark mixing, Cabibbo-Kobayashi-Maskawa (CKM), matrix. These phenomena offer multiple experimental opportunities to probe the masses and mixing of the extra quarks; for a summary of relevant constraints see \cite{Alves:2023ufm}. 

Regarding the strong CP problem, the Peccei-Quinn (PQ) \cite{Peccei:1977hh, Peccei:1977ur} and the Nelson-Barr \cite{Nelson:1983zb, Barr:1984qx} mechanisms provide two fundamentally distinct yet effective solutions. 
In the Nelson-Barr scheme, one assumes that 
CP is a fundamental symmetry of nature that is spontaneously broken, and the couplings of the SM to the CP breaking sector are arranged to guarantee vanishing $\btheta$ at tree level.
The challenge is to generate an order one CKM phase 
and, simultaneously, suppress corrections to $\btheta$ sufficiently.
Corrections may already show up at one-loop\,\cite{Bento:1991ez}, although it may be postponed to two-loops by using a nonconventional CP symmetry\,\cite{Cherchiglia:2019gll}.
Quite similarly to the axion solution, the Nelson-Barr solution also suffers from a quality problem\,\cite{Dine:2015jga,Asadi:2022vys,Choi:1992xp}, which can be ameliorated by additional gauge symmetries\,\cite{Asadi:2022vys}, supersymmetry \,\cite{Dine:2015jga} or strong dynamics\,\cite{Valenti:2021xjp}.

In Nelson-Barr settings, the spontaneous breaking of CP is transmitted to the SM through the mixing of SM quarks with heavy vector-like quarks of Nelson-Barr type (NB-VLQs)\,\cite{Cherchiglia:2020kut,Cherchiglia:2021vhe,Alves:2023cmw,Valenti:2021rdu}.
The NB-VLQs that mediate the CP breaking may lie at much lower energies, only constrained to be above the TeV scale from collider searches.
To comply with the Barr criteria \cite{Barr:1984qx}, the NB-VLQs can only be electroweak singlets or doublets, in the same representations of SM quarks.
The case of doublets was argued to lead to too large corrections to  $\btheta$ \,\cite{Vecchi:2014hpa}.
For singlets, we have found for one\,\cite{Cherchiglia:2020kut,Cherchiglia:2021vhe} or two\,\cite{Alves:2023cmw}
NB-VLQs that they typically couple to the SM quarks and Higgs following the hierarchy of the CKM last row or column, a feature that alleviates the strongest flavour constraints that apply to the first two quark families.
For special points for one NB-VLQ, one can even switch off the coupling of such a heavy quark with a given quark flavour\,\cite{Alves:2023cmw,Cherchiglia:2021syq}.

Several connections between the strong CP and the fermion mass hierarchy problems have been explored in the literature.  There are proposals with extra quarks getting massive at the PQ scale and leading to seesaw-suppressed masses for all the standard quarks \cite{Kang:1985ua}, or tree-level (seesaw) and loop-suppressed masses for the different generations \cite{Babu:1988fn}. Interrelated solutions to these problems have also been investigated in SM gauge extensions featuring the PQ mechanism, such as a $SU(3)_H$ extension \cite{Berezhiani:1990wn}, a 3-3-1 model \cite{Dias:2018ddy} and a non-universal $U(1)_X$ model \cite{Garnica:2019hvn}. Furthermore, the $\btheta$ phase can be generated at the two-loop level in extended left-right models incorporating seesaw mechanisms for the SM fermions ~\cite{Babu:1988mw,Babu:1989rb}. On the other hand, the Froggat-Nielsen and the Peccei-Quinn mechanisms can be realised synergistically by identifying the Froggat-Nielsen flavon with the PQ axion \cite{Ema:2016ops,Calibbi:2016hwq}.  

In this work, we propose a new framework where solutions to both the strong CP problem and the flavour puzzle are also directly linked. 
We focus on the Nelson-Barr solution to the strong CP problem and explore the possibility that the newly introduced VLQs generate seesaw-suppressed masses for the SM quarks.

The outline of this article is the following:
In Sec.\,\ref{sec:NB}, we present an overview of the NB-VLQs that transmit CP violation to the SM in the Nelson-Barr mechanism. The ``seesaw limit'' of the Nelson-Barr mechanism is identified and important relations are derived in Sec.\,\ref{sec:NBSS}. In the following section, Sec.\,\ref{sec:NBSSforD}, we study a scenario where three down-type vector-like quarks are added to the SM while in Sec.\,\ref{sec:NBSSforU}, we explore the consequences of the SM extended by two up-type vector-like quarks. Finally, our conclusions are summarised in Sec.\,\ref{sec:Con}.

\section{The Nelson-Barr mechanism}\label{sec:NB}

In this section, we provide an overview of the so-called Nelson-Barr mechanism \cite{Nelson:1983zb, Barr:1984qx}, which tackles the strong CP problem via the spontaneous breaking of the CP symmetry. For simplicity and clarity, we focus here on the case in which only the down-type quark sector of the SM is extended. Similar steps can be followed when extending the up-type or both quark sectors instead. We start by introducing a set of $n$ vector-like quarks (VLQs) transforming under the SM symmetry group, $\mathcal{G}_{SM}=SU(3)_C\times SU(2)_L \times U(1)_Y$, as 
\begin{equation}\label{eq:BLR}
B_{aL,R}\sim (\mathbf{3},\mathbf{1},-1/3), \quad \mbox{with} \quad a=1,\cdots,n\,.
\end{equation}
In addition to $\mathcal{G}_{SM}$, two discrete symmetries, $CP$ and $\ZZ_2$, are imposed. Under the latter, only the new quarks transform non-trivially. 

Given these ingredients, the Yukawa interactions for the quarks can then be written as
\eqali{
\label{eq:Yukq}
-\lag_q=\bar{q}_{iL}\cY^u_{ij} \tilde{H}u_{jR}
+\bar{q}_{iL}\cY^d_{ij} Hd_{jR}+\bar{B}_{aL}\cM^B_{ab} B_{bR}+
\bar{B}_{aL}\cM^{Bd}_{aj} d_{jR}+h.c.,
}
with the last term, which softly breaks $\ZZ_2$ as well as $CP$, arising from spontaneous symmetry breaking when scalar fields -- singlets under $\mathcal{G}_{SM}$ -- acquire non-trivial complex vevs\footnote{For instance, we can have $\mathcal{M}^{Bd} = \mathcal \sum_j\mathcal{Y}^{Bd}_j\langle\varphi_j\rangle$, where $\mathcal{Y}^{Bd}_j$ are ($n\times3$) real matrices and $\langle\varphi_j\rangle = v_j e^{i \alpha_j}$.}.  All the other terms are assumed to conserve $CP$ (as well as $\ZZ_2$), implying that $\mathcal{Y}^u$ and $\mathcal{Y}^d$ -- both $3\times 3$ -- as well as $\mathcal{M}^B$ -- $n\times n$ -- are real matrices. Once the Higgs doublet acquires a vev, $\langle H\rangle = (0, v/\sqrt{2})^T$, we can write down the following $(3+n)\times (3+n)$ mass matrix for the down-type quarks,\footnote{%
Note that in generic settings some zeros may be produced by weak basis changes; see a discussion in Ref.\,\cite{Alves:2023ufm}. Here, due to the softly broken $CP$ and $\ZZ_2$ symmetries, this structure 
with real entries
cannot result from a weak basis change.
}
\begin{eqnarray}\label{eq:MdB}
    \cM^{d+B} = \begin{pmatrix}
        \frac{v}{\sqrt{2}} \mathcal{Y}^d_{(3\times 3)} & 0_{(3\times n)} \\
        \mathcal{M}^{Bd}_{(n\times 3)} & \mathcal{M}^{B}_{(n\times n)}        
\end{pmatrix}\,,
\end{eqnarray}
whereas for the up-type quarks, we have $m^u =\frac{v}{\sqrt{2}} \mathcal{Y}^u_{(3\times 3)}$, which is taken diagonal for simplicity.

The contributions at tree level to $\bar{\theta}$ coming from the quark sector ($\theta_q$) vanish because $\arg(\det \cM^{d+B}) = \arg(\det m^u) =0$. Moreover, since the $CP$ symmetry is only broken (spontaneously) through $\mathcal{M}^{Bd}$, the bare $\theta_{QCD}$ term also vanishes. Consequently, the effective parameter $\bar{\theta} = \theta_{QCD}+\theta_{q}$ vanishes at tree level. 
The leading one-loop corrections can also be made small by, e.g., suppressing the Yukawa couplings of the VLQs with the CP-breaking scalars. There are, however, irreducible contributions at 3-loops\,\cite{Valenti:2021rdu} which might be more important.
For generic points, these contributions are relevant for two or more up-type VLQs but they are still compatible with current bounds\,\cite{Alves:2023cmw}.
Their importance in the \textit{seesaw limit} of Sec.\,\ref{sec:NBSS} is discussed in Secs.\,\ref{sec:NBSSforD} and \ref{sec:NBSSforU}.

It is convenient to perform a unitary transformation on the right-handed fields $U_R$, so that $\mathcal{M}^{d+B}$ becomes \cite{Alves:2023cmw, Alves:2023ufm}
\eq{
\label{eq:MWB}
M^{d+B} \equiv \mathcal{M}^{d+B}\, U_R = \begin{pmatrix}
        \frac{v}{\sqrt{2}} Y^d_{(3\times 3)} & \frac{v}{\sqrt{2}} Y^B_{(3\times n)} \\
        0_{(n\times 3)} & M^{B}_{(n\times n)}        
\end{pmatrix}\,,
}
where
\eq{
\label{eq:UR}
U_R = 
\begin{pmatrix}
        \id_3 & \tilde{w}_R \\
        -\tilde{w}_R^\dagger & \id_n        
\end{pmatrix}  \begin{pmatrix}
        \left(\id_3+\tilde{w}_R\tilde{w}_R^\dagger\right)^{-1/2} & 0 \\
        0 & \left(\id_n+\tilde{w}_R^\dagger \tilde{w}_R\right)^{-1/2}  
\end{pmatrix}\,.
}
These mass matrices are then related via
\begin{equation} 
\label{eq:WBrels}
\begin{aligned}
Y^d &= \mathcal{Y}^d(\id_3+\tilde{w}_R\tilde{w}_R^\dagger)^{-1/2}\,,\\
Y^B &= \mathcal{Y}^d\tilde{w}_R(\id_n+\tilde{w}_R^\dagger\tilde{w}_R)^{-1/2}\,,\\
M^B &= \mathcal{M}^B(\id_n+\tilde{w}_R^\dagger\tilde{w}_R)^{+1/2}\,,
\end{aligned}
\end{equation}
with
\eq{
\tilde{w}_R^\dagger = {\mathcal{M}^B}^{-1}\mathcal{M}^{Bd}\,.
}
In addition, by using the relation $\tilde{w}_R(\id_n+\tilde{w}_R^\dagger\tilde{w}_R)^{-1/2}=(\id_3+\tilde{w}_R\tilde{w}_R^\dagger)^{-1/2}\tilde{w}_R$, we see that the Yukawa coupling of the heavy quarks and Yukawa coupling of the SM quarks obey an interesting relation\,\cite{Cherchiglia:2020kut,Alves:2023cmw}: 
\eq{
\label{YB:Yd}
Y^B = Y^d\, \tilde{w}_R\,.
}
It follows that $Y^B$ roughly follows the hierarchy of the CKM entries. 

Coming back to $\mathcal{M}^{d+B}$ in Eq.\,(\ref{eq:MdB}), we notice that the terms $\mathcal{M}^{Bd}$ and $\mathcal{M}^{B}$ conserve the SM symmetry structure, and, as such, we can safely assume that they emerge at much higher energy scales than $v$, which simply translates into 
\begin{equation}\label{eq:lim1}
    S\equiv \mathcal{M}^{Bd} {\mathcal{M}^{Bd}}^\dagger + \mathcal{M}^B {\mathcal{M}^{B}}^T \gg v^2/2 \left(\mathcal{Y}^d{\mathcal{Y}^d}^T\right).
\end{equation} 
In this case, we can block diagonalise $\mathcal{H}= \cM^{d+B}{\cM^{d+B}}^\dagger$ via the unitary transformation on the left-handed fields 
\begin{equation}\label{eq:Hdiag}
U_L^\dagger\, \mathcal{H}\, U_L \simeq \begin{pmatrix}
    \mathcal{H}_d & 0 \\
    0 & \mathcal{H}_B
\end{pmatrix}\,,
\end{equation}
where $U_L$ is a function of $\tilde{w}_L$ following the same ansatz for $U_R$ in Eq.\,(\ref{eq:UR}). At leading order, we obtain
\begin{eqnarray}
\label{eq:Hd}
\mathcal{H}_d&\simeq&\frac{v^2}{2}\cY^d
(\id_3-{\mathcal{M}^{Bd}}^{\dagger}S^{-1}\mathcal{M}^{Bd}){\cY^d}^\tp
\,,\cr
\mathcal{H}_B&\simeq&  \mathcal{M}^{Bd} {\mathcal{M}^{Bd}}^\dagger + \mathcal{M}^B {\mathcal{M}^{B}}^T
\,= S\,, \\
\tilde{w}_L&\simeq& \frac{v}{\sqrt{2}}\,\mathcal{Y}^d\,\mathcal{M}^{Bd\, \dagger}\, S^{-1}\,.\nonumber
\end{eqnarray}
Finally, by further diagonalising $\mathcal{H}_B$ by means of a $n\times n$ unitary transformation, we find the (squared) masses of the heavy, non-standard quarks. On the other hand, $\mathcal{H}_d$ is approximately diagonalised by the CKM matrix so that $V_{CKM}^\dagger \mathcal{H}_d V_{CKM} = \diag(m_d^2,m_s^2,m_b^2)$, where $m_{u,s,b}$ are the masses of the down, strange and bottom quarks, respectively.
For up-type VLQs, we should replace $V_{CKM}\to V_{CKM}^\dag$.

\section{The seesaw limit of the Nelson-Barr mechanism}\label{sec:NBSS}

We now explore the {\it seesaw limit} of the model described in the previous section. For simplicity, we first assume that the vector-like quarks appear in the same number as the standard quarks, {\it i.e.} $n=3$. In this case, the down-type quark mass matrix in Eq.\,(\ref{eq:MdB}) becomes\footnote{Notice that by swapping the first and second block columns, which is equivalent to a change of basis, the resulting matrix has a type-I Dirac-seesaw texture $
    \tilde{\cM}^{d+B} = \begin{pmatrix}
         0 & \frac{v}{\sqrt{2}} \mathcal{Y}^d  \\
         \mathcal{M}^{B} & \mathcal{M}^{Bd}        
\end{pmatrix}\,.$}
\begin{eqnarray}\label{eq:MD3x3}
    \cM^{d+B} = \begin{pmatrix}
        \frac{v}{\sqrt{2}} \mathcal{Y}^d_{(3\times 3)} & 0_{(3\times 3)} \\
        \mathcal{M}^{Bd}_{(3\times 3)} & \mathcal{M}^{B}_{(3\times 3)}        
\end{pmatrix}\,.
\end{eqnarray}
The seesaw limit consists in the special case of the condition Eq.\,\eqref{eq:lim1} in which\,\footnote{%
We distinguish between the condition Eq.\,\eqref{eq:lim1} for the validity of the {\it seesaw approximation} for the mixing of the left-handed fields and this {\it subcase} where the term that softly breaks CP is dominant.
}
\begin{equation}\label{eq:sslim}
    {\mathcal{M}^{Bd}}{\mathcal{M}^{Bd}}^\dagger \gg \mathcal{M}^{B}{\mathcal{M}^{B}}^T .
\end{equation}
This limit allows for the expansion of ${\mathcal{M}^{Bd}}^{\dagger}S^{-1}\mathcal{M}^{Bd}$  in Eq.\,(\ref{eq:Hd}) as
\eqali{\label{eq:wexp}
% ww^\dag&=
&{\cM^{Bd}}^\dag \Big[\cM^{Bd}{\cM^{Bd}}^\dag +\cM^B{\cM^B}^T\Big]^{-1}\cM^{Bd}
\cr
&=
{\cM^{Bd}}^\dag \Big[\id_n + \left(\cM^{Bd}{\cM^{Bd}}^\dag\right)^{-1}\cM^B{\cM^B}^T\Big]^{-1} \left(\cM^{Bd}{\cM^{Bd}}^\dag\right)^{-1}\cM^{Bd}
\cr
&\simeq
{\cM^{Bd}}^\dag \left(\cM^{Bd}{\cM^{Bd}}^\dag\right)^{-1}\cM^{Bd} -{\cM^{Bd}}^\dag \left(\cM^{Bd}{\cM^{Bd}}^\dag\right)^{-1}\cM^B{\cM^B}^T\left(\cM^{Bd}{\cM^{Bd}}^\dag\right)^{-1}\cM^{Bd}
\cr
&\simeq
\id_3-\left({\cM^{Bd}}^{-1}\cM^{B}\right)\left({\cM^{Bd}}^{-1}\cM^{B}\right)^\dag\,,
}
where the last step is only possible when $\mathcal{M}^{Bd}$ is a square matrix, while the second line is only possible for $n=1,2,3$. 
Substituting this result back into Eq.\,(\ref{eq:Hd}), we obtain
\begin{equation}\label{eq:HdSS}
    \mathcal{H}_d\simeq\left(\frac{v}{\sqrt{2}}\cY^d{\cM^{Bd}}^{-1}\cM^{B}\right)\left(\frac{v}{\sqrt{2}}\cY^d{\cM^{Bd}}^{-1}\cM^{B}\right)^\dagger.
\end{equation}
Therefore, we find that the standard quark masses are seesaw suppressed with respect to the electroweak scale
$v$ by a factor 
\eq{
\label{n=3:epsilon}
\epsilon \equiv \tilde{w}_R^{\dag -1} ={\cM^{Bd}}^{-1}\cM^{B}\,,
}
which has entries much smaller than unity.
This quantity is simply $\tilde{w}^{\dag -1}$ in the notation of Ref.\,\cite{Alves:2023cmw}.

If, for instance, we assume that $\mathcal{O}(\epsilon) =10^{-2}$, then the quark masses are roughly given by $m^d\simeq \frac{v}{\sqrt{2}}\cY^d\times 10^{-2}$. Thus, instead of being proportional to $v\sim \mathcal{O}(10^2)$ GeV, down-type quark masses are now proportional to an effective scale $v\times \epsilon\sim O(1)$ GeV. Consequently, the CP-conserving Yukawa entries in $\mathcal{Y}^d$ can be two orders of magnitude larger than the ones in the Standard Model.

Furthermore, to study the physical cases consistent with the current quark mixing picture, it is convenient to solve Eq.\,\eqref{eq:HdSS} for $\cM^{Bd}$ in terms of the other matrices as
\eq{\label{eq:MBdCI}
\cM^{Bd}=\cM^B U(\hY^d)^{-1}V_{\rm CKM}^\dag\cY^d\,,
}
where $\hY^d = \mathrm{diag}(y_d,y_s,y_b)$ is the diagonal Yukawa coupling matrix for the down-type quarks in the SM, and $U$ is a generic $3\times 3$ unitary matrix. 
This relation has been obtained by assuming that $\mathcal{H}^d$ is diagonalised by the CKM matrix, {\it i.e.} $V_{CKM}^\dagger \mathcal{H}_d V_{CKM} \simeq \mathrm{diag}(m_d^2,m_s^2,m_b^2)$; therefore, all points generated from Eq. (\ref{eq:MBdCI}) are automatically in agreement with the observed values of the CKM matrix parameters.
This inversion is particularly simple in the seesaw limit; away from this limit a more complicated parametrization similar to the ones in Refs.\,\cite{Cherchiglia:2020kut,Alves:2023cmw} will be needed.
Without loss of generality, we can choose a basis in which $\cM^B$ is diagonal and $\cY^d=O_{d_L}\hat{\cY}^d$, with orthogonal $O_{d_L}$ and diagonal $\hat{\cY}^d$. 
Finally, we can see that if $\cY^d$ has order-one entries, $\cM^{Bd}$ is larger than $\cM^B$ by, at least, $1/y_b\sim 10^2$. This is clear if we substitute Eq.\,\eqref{eq:MBdCI} into Eq.\,\eqref{n=3:epsilon}, obtaining
\eq{
\label{eps:n=3:yuk}
\epsilon={\cY^d}^{-1}V_{\rm CKM} \hY^d U^\dag\,.
}
Note that $V_{\rm CKM}$ is the CKM matrix of the SM with two additional phases\,\cite{Cherchiglia:2020kut}:
\eq{
V_{\rm CKM}=\diag(1,e^{i\beta_1},e^{i\beta_2})V_{0}\,,
}
where $V_0$ is, e.g., the standard parametrization.
These additional phases arise because the Yukawa $Y^d$ in Eq.\,\eqref{eq:WBrels} loses rephasing freedom from the left to keep $\cY^d$ real. Notice that, in the particular case that the generic matrices $U$ and $O_{d_L}$ are diagonal, only $\cM^{Bd}$ and $\eps$ are nondiagonal owing to the CKM matrix. As in the SM, we do not aim to explain the origin of the hierarchy exhibited by the CKM but rather provide a solution for the hierarchy among the quark masses. In other words, the entries of $\hat{\cY}^d$ can be much larger than the down Yukawa matrix in the SM, while their mixing comes from phenomenological observation.

Turning to the heavy quarks, at leading order, their mass matrix is given by
\eq{
\label{def:HB}
\mathcal{H}_B=\cM^B{\cM^B}^T + \cM^{Bd}{\cM^{Bd}}^\dag\,.
}
In the seesaw limit, this mass matrix is dominated by $\cM^{Bd}$ which can be related to $\cM^B$ by
\eq{
\label{MBd:eps} 
\cM^{Bd}=\cM^{B}\epsilon^{-1}\,.
}
Additionally, the leading contribution to the mixing parameter $\tilde{\omega}_L$ in Eq.\,(\ref{eq:Hd}) is given by
\eq{
\label{eq:Fn=3}
\tilde{\omega}_L\simeq \frac{v}{\sqrt{2}}\,\mathcal{Y}^d\,{\mathcal{M}^{Bd}}^{-1}\,.
}
Finally, the relation \eqref{YB:Yd} becomes 
\eq{
\label{YB}
Y^B=Y^d\epsilon^{\dag -1}\,.
}
This relation now indicates that $Y^B$ is larger than the SM Yukawa $Y^d$ by the factor $\epsilon^{\dag -1}$.

\subsection*{Special cases: $n<3$, incomplete seesaw}

The case presented above assumes the introduction of a vector-like quark per standard quark generation, {\it i.e.}, $n=3$. As a result, the seesaw mechanism suppresses the overall mass scale for all the standard quarks. Nevertheless, the seesaw limit is of interest even in cases with fewer vector-like quarks, $n<3$. This is particularly true for the up-type sector, as we will discuss in Sec. \ref{sec:NBSSforU}. 

In scenarios with fewer vector-like quarks than standard ones ($n=1,2$), the matrix $\mathcal{M}^{Bd}_{(n\times3)}$ is not square and, as such, the approximation in Eq.\,(\ref{eq:wexp}) is only valid up to the second to last line, {\it i.e.},
\eqali{\label{eq:wexp2}
{\mathcal{M}^{Bd}}^{\dagger}S^{-1}\mathcal{M}^{Bd}
\simeq &\,\,
{\cM^{Bd}}^\dag \left(\cM^{Bd}{\cM^{Bd}}^\dag\right)^{-1}\cM^{Bd}\cr
&-{\cM^{Bd}}^\dag \left(\cM^{Bd}{\cM^{Bd}}^\dag\right)^{-1}\cM^B{\cM^B}^T\left(\cM^{Bd}{\cM^{Bd}}^\dag\right)^{-1}\cM^{Bd}
\,.
}
Thus, in contrast with the case where $n=3$, not all standard quark masses, obtained from Eq.\,(\ref{eq:Hd}) with Eq.\,(\ref{eq:wexp2}), become suppressed. In this case, only $n$ of the standard quarks get seesaw-suppressed masses, while the remaining $(3-n)$ masses remain proportional to the Higgs vev $v$.

To understand this, notice that the first term in Eq.\,\eqref{eq:wexp2}, that is, the $3 \times 3$ 
matrix $P={\cM^{Bd}}^\dag \left(\cM^{Bd}{\cM^{Bd}}^\dag\right)^{-1}\cM^{Bd}$ has rank $n$ with the non-vanishing eigenvalues equal to unity, since it is a projection matrix obeying $P^2=P$ and 
$\cM^{Bd}$ is rank $n$.
Therefore, taking, for instance, $n=2$ and rotating the fields to a basis where $P$ is diagonal, we have
\begin{eqnarray}
\id_3-{\mathcal{M}^{Bd}}^{\dagger}S^{-1}\mathcal{M}^{Bd} = \id_3 - \mathrm{diag}(1,1,0) + \mathcal{O}(\delta) = \mathrm{diag}(0,0,1) + \mathcal{O}(\delta)\,,  
\end{eqnarray}
where $\mathcal{O}(\delta)\ll 1$ represents the seesaw suppressed terms in ${\mathcal{M}^{Bd}}^{\dagger}S^{-1}\mathcal{M}^{Bd}$ -- the (rotated) second term in Eq.\,(\ref{eq:wexp2}). As $\mathcal{H}_d = (v^2/2){\mathcal{Y}^d}(\id_3-{\mathcal{M}^{Bd}}^{\dagger}S^{-1}\mathcal{M}^{Bd}){\mathcal{Y}^d}^T$, {\it cf.}\ Eq.\,\eqref{eq:Hd}, 
for a $\cY^d$ with similar entries, two of the quark masses are seesaw suppressed, while one remains non-suppressed. Similarly, when $n=1$, one quark mass becomes seesaw suppressed and the other two do not.

In such cases, instead of Eq.\,\eqref{eq:MBdCI}, we can use the formulae derived in Refs. \cite{Cherchiglia:2020kut,Alves:2023cmw} to find benchmarks satisfying the current quark mixing picture. 
In particular, the CKM matrix is automatically taken into account.
The exact parametrization involves solving the first equation of \eqref{eq:Hd} for $\cY^d$.
To that end, we first rewrite it as
\eq{
\label{inversion:1}
{\cY^d}^{-1}\frac{2}{v^2}\mathcal{H}_d{\cY^d}^{\tp -1}
=\id_3-{\mathcal{M}^{Bd}}^{\dagger}S^{-1}\mathcal{M}^{Bd}
\equiv \Omega\,.
}
If we split $\Omega=\Omega_1+i\Omega_2$ in terms of its real and imaginary parts, we can solve the real part of \eqref{inversion:1} by
\eq{
\label{cY:inversion}
\cY^d=\bigg[\frac{2}{v^2}\re\mathcal{H}_d\bigg]^{1/2}\mathcal{O}\Omega_1^{-1/2}\,,
}
where $\mathcal{O}$ is a real orthogonal matrix to be determined from the imaginary part of \eqref{inversion:1}, i.e., 
\eq{
\Omega_1^{-1/2}\Omega_2\Omega_1^{-1/2}
=\mathcal{O}^\tp [\re\mathcal{H}_d]^{-1/2}\im(\mathcal{H}_d)[\re\mathcal{H}_d]^{-1/2}\mathcal{O}\,.
}
By choosing an appropriate basis for $\Omega$, a parametrization can be found for $n=1$\,\cite{Cherchiglia:2020kut} and $n=2$\,\cite{Alves:2023cmw}.
For the latter, additional parameters are necessary to describe $\cM^{Bd}$ and $\cM^B$.

\section{Nelson-Barr seesaw for the down-type quarks}\label{sec:NBSSforD}

We now discuss in more detail the consequences of having three down-type VLQs participating in the Nelson-Barr mechanism as introduced in Sec.\,\ref{sec:NBSS}.
We generate points by using Eq.\,\eqref{eq:MBdCI} in the basis where
\eq{
\cM^B=\text{diagonal}\,,\quad
\cY^d=O_{d_L}\hat{\cY}^d\,,
}
where $O_{d_L}$ is a real orthogonal matrix, and $\hat{\cY}^d$ is the diagonal matrix of singular values.
In the analyses that follow, we vary the diagonal entries of $\cM^B$ in the range $[10,10^4]$ GeV and $\theta_{d_L}^i\in [10^{-4},2\pi]$, where $\theta_{d_L}^i$ are the three angles that parametrise $O_{d_L} \equiv O_{d_L}(\theta_{d_L}^i)$. The SM quark masses and CKM matrix parameters are fixed to their best-fit values \cite{ParticleDataGroup:2022pth}. Furthermore, we take into account bounds from VLQ searches\,\cite{ATLAS:2011tvb} to filter out solutions with one or more VLQs lighter than $1.3$ TeV.

\begin{figure}[h!]
\centering
\includegraphics[scale=0.7]{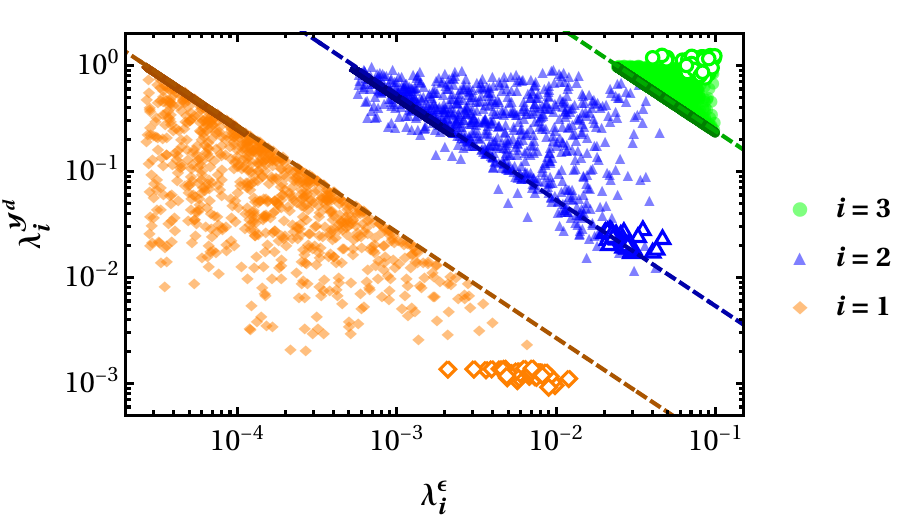}
\caption{Singular values $\lambda_i^X$ of $X=\cY^d$ as a function of singular values for $X=\eps$.
The filled (hollow) points denote the generic (hierarchical) $\cY^d$ case described in Eq.\,\eqref{calY:gen-hier}. Diagonal lines represent $\lambda_i^{\mathcal{Y}^d}\lambda_i^{\epsilon} = y_i$ (no sum), where $y_i$ are the SM Yukawa couplings. The darker points comply with $\mathcal{Y}^d = y\,\id_3$; see text for more details.
}
\label{fig:calY-epsi}
\end{figure}

Let us first analyze our seesaw expansion matrix $\epsilon$ in Eq.\,\eqref{n=3:epsilon} which is the ratio between the CP-conserving bare VLQ mass and the dominant CP-violating contribution.
In the seesaw limit, this matrix should be naturally small as it gives in Eq.\,\eqref{eq:HdSS} the suppression of $v\cY^d$ to generate the SM down quark masses.
Another expression of this feature can be seen in Eq.\,\eqref{eps:n=3:yuk}. To ensure that the seesaw limit, see Eq.\,(\ref{eq:sslim}), is satisfied by all of our solutions, we select points for which all of the singular values of $\cM^{Bd}$ are at least one order of magnitude larger than the singular values of $\cM^B$ or, equivalently, all of the eigenvalues of $\sqrt{\cM^{Bd}\cM^{Bd\,\dagger}}$ are at least one order of magnitude larger than those of $\sqrt{\cM^B\cM^{B\,\dagger}}$.
This suppression can be seen in Fig.\,\ref{fig:calY-epsi} where we show the singular values of $\cY^d$ as a function of the singular values of $\epsilon$.
In the plot, we denote as $\lambda_i^{X}$ each singular value of the matrix $X$ in increasing order.
The filled and hollow symbols denote two sets of points:
\eqali{
\label{calY:gen-hier}
\text{Generic $\cY^d$ (filled):}&\quad \hat{\cY}^d_{ii}\in [10^{-3},1];
\cr
\text{Hierarchical $\cY^d$ (hollow):}&\quad \frac{y_b}{y_i}\hat{\cY}^d_{ii} \in [0.75,1.25].
}
In the second case, the singular values $\hat{\cY}^d_{ii}$ follow the hierarchy of the SM Yukawas $y_i=(y_d,y_s,y_b)$ allowing for a variation of 25\%.
We can see in the figure in the generic $\cY^d$ set that $\cY^d$ of the same order requires hierarchical $\epsilon$ to account for the hierarchy of the SM Yukawa $Y^d$.
On the other hand, as $\cY^d$ becomes hierarchical, a less hierarchical $\epsilon$ is necessary.
In the extreme case of $\cY^d$ following the hierarchy of the SM Yukawa (hierarchical $\cY^d$ set), non-hierarchical values for $\epsilon$ are allowed. 
Finally, one could wonder if the observed features in Fig.~\ref{fig:calY-epsi} could be due to some hidden hierarchy in the matrices $U$ and $O_{d_L}$. 
We have checked that, even in the case that they are equal to the identity matrix, the overall behaviour is maintained, i.e., the hierarchies of the down Yukawas can be attributed to $\cY^d$, to $\eps$ or to an interplay between them.

Moreover, if we also require that $\mathcal{Y}^d = y\,\id_3$, we obtain the darker points shown in Fig.~\ref{fig:calY-epsi}, where solutions are attained in the region $y>10^{-1}$. In this scenario, it is clear that the observed hierarchies among the down-type quark masses can be entirely attributed to hierarchies in $\eps$, instead of $\cY^d$.

\begin{figure}[h!]
\centering
\includegraphics[scale=0.7]{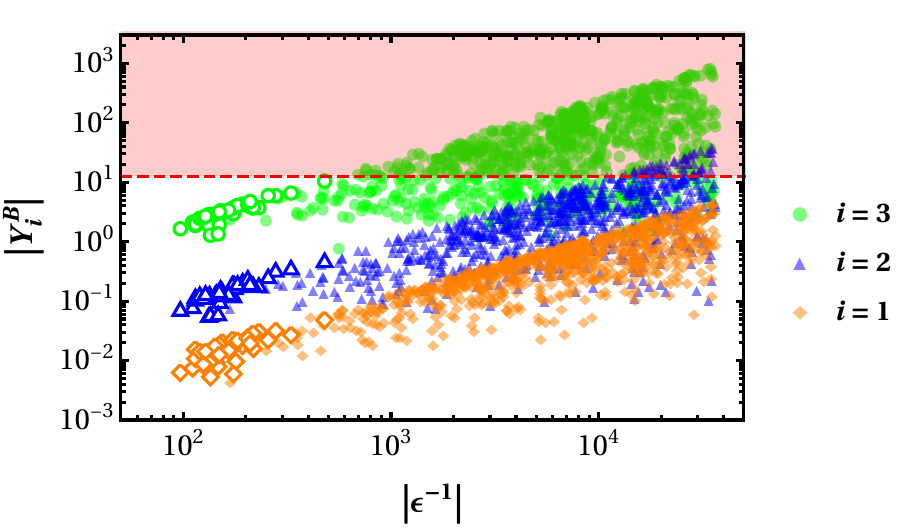}
\caption{
The norms $|Y^B_i|$ of VLQ Yukawa couplings in Eq.\,\eqref{def:YBi} as a function of the norm of $\eps^{-1}$ in Eq.\,\eqref{norm:eps-1}.
The filled (hollow) points denote the generic (hierarchical) $\cY^d$ case described in Eq.\,\eqref{calY:gen-hier}.
The red-shaded region denotes the perturbativity exclusion $|Y^B_i|<4\pi$.
}
\label{fig:YB-eps}
\end{figure}

We can now illustrate the behaviour of the Yukawa couplings $Y^B_{ia}$, cf. Eq.\,\eqref{YB}, of the heavy quark $B_{aR}$ to the SM Higgs and quark $q_{iL}$.
Instead of the individual couplings, we take the norm of the couplings to each SM family:
\eq{
\label{def:YBi}
|Y^B_{i}| \equiv \sqrt{\sum_{a=1}^3 \big|Y^B_{ia}\big|^2}\,.
}
In Fig.\,\ref{fig:YB-eps} we show this norm of Yukawas as a function of the norm of $\eps^{-1}$:
\eq{
\label{norm:eps-1}
|\epsilon^{-1}| \equiv \sqrt{\operatorname{tr}[\epsilon^{-1}\epsilon^{-1\dag}}]\,.
}
We can see that the rough hierarchy follows
\eq{
\label{typical.YB}
|Y^B_1|:|Y^B_2|:|Y^B_3|\sim |V_{ub}|:|V_{cb}|:|V_{tb}|
\sim 0.0036:0.04:1
\,,
}
analogous to the case of one or two VLQs\,\cite{Cherchiglia:2020kut,Alves:2023cmw}.
We also observe that generic $\cY^d$ has larger Yukawas and larger $|\eps^{-1}|$ while the hierarchical $\cY^d$ case has smaller Yukawas and smaller $|\eps^{-1}|$ in accordance to Eq.\,\eqref{YB}.
For the former set, many points are excluded by the perturbativity 
constraint $|Y^B_i|<4\pi$ which is shaded red.

\begin{figure}[h!]
\centering
\includegraphics[scale=0.7]{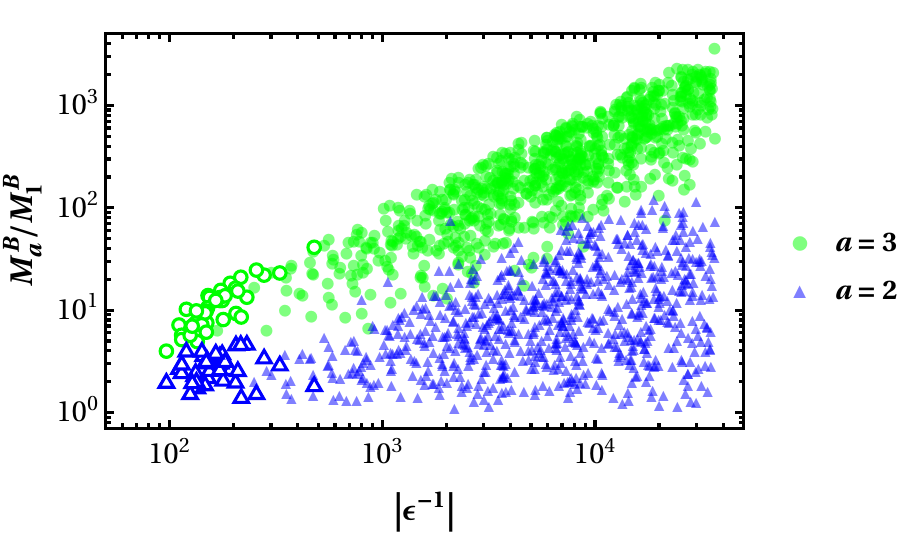}
\caption{
The ratio $M^B_a/M^B_1$ of the heavy quarks as a function of the norm of $\eps^{-1}$ in Eq.\,\eqref{norm:eps-1}.
}
\label{fig:MB-eps}
\end{figure}

To get a sense of the spectrum for the VLQs, we show in Fig.\,\ref{fig:MB-eps} the ratio of the heavy masses in Eq.\,\eqref{def:HB} compared to their lightest mass as a function of $|\eps^{-1}|$.
In accordance with Eq.\,\eqref{MBd:eps}, we see that larger $|\eps^{-1}|$ tends to lead to a larger hierarchy among the heavy quarks.
For the hierarchical $\cY^d$ set where the entries of $\eps$ are all of the same order, the heavy masses are also non-hierarchical.

\begin{figure}[h!]
\centering
\includegraphics[scale=0.9]{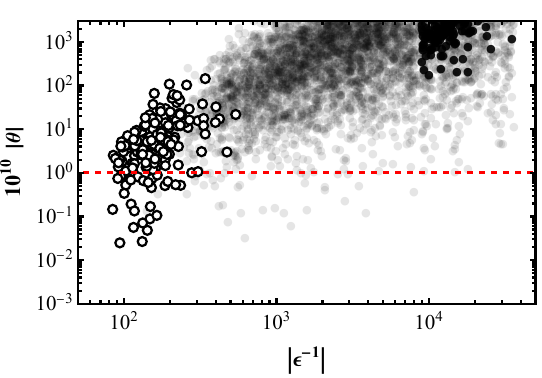}
\caption{
Estimate \eqref{invariants:n=2:down} of the irreducible 3-loop contributions to $\btheta$ against $|\eps^{-1}|$ in Eq.\,\eqref{norm:eps-1} for three down-type VLQs in the seesaw limit.
The differences among hollow, filled and dark points follow the convention of Fig.\,\ref{fig:calY-epsi}.
}
\label{fig:theta-eps:3vlq}
\end{figure}

Finally, we show in Fig.\,\ref{fig:theta-eps:3vlq} the estimate to the 
3-loop irreducible contribution to $\btheta$ coming from VLQ exchange against $|\eps^{-1}|$.
The dominant contribution is expected to be\,\cite{Valenti:2021xjp}
\eqali{
\label{invariants:n=2:down}
\text{down-type $n_B\ge 2$: }\quad
\left(\frac{1}{16 \pi^2}\right)^3\im \text{Tr}\left( \left[{Y^B}^\dag Y^u{Y^{u}}^\dagger {Y^B}, {Y^B}^\dag{Y^B}\right]F\left({M^B}^\dagger M^B\right)\right),}
with model dependent order one coefficient, and we define $F(A)=A/\operatorname{Tr}[A]$.
The distinction among hollow, filled and dark (opaque) points follows the convention of Fig.\,\ref{fig:calY-epsi}. We, however, have already filtered out the points that violate the perturbativity of the Yukawas.
We can see that the dark points, corresponding to the case where $\cY^d\propto \id_3$ carries no hierarchy, are severely excluded by the limits on $\btheta$ from the neutron electric dipole moment.
The other subsets, although severely constrained, contain points that pass the limit.
In particular, the case of hollow points, cf.\,\eqref{calY:gen-hier}, where $\cY^d$ carries the hierarchy among the SM Yukawas, contains a reasonable quantity of viable points.

\section{Nelson-Barr seesaw for the up-type quarks}\label{sec:NBSSforU}

We now turn to the up-type quarks. Since the top Yukawa is already of order one, we aim to provide an explanation for the observed hierarchy among up and charm quarks only. This can be achieved by introducing two NB-VLQs. In this case, we cannot use Eq.\,\eqref{eq:MBdCI} since it was derived under the assumption that $\cM^{Tu}$ is invertible (it is a $2\times3$ matrix now)\footnote{For up-type quarks, instead of $B$ and $d$, vector-like and standard quarks are denoted by $T$ and $u$, respectively.}. We instead rely on a parametrization devised in~\cite{Cherchiglia:2020kut,Alves:2023cmw}, which takes the known up-type quark masses, CKM mixing and the lightest VLQ mass as inputs (among other non-physical variables). We fixed the SM quantities to their best-fit values \cite{ParticleDataGroup:2022pth}, the lightest VLQ mass to 1.3 TeV~\cite{Alves:2023ufm} while the other parameters were varied in a broad range. Notice that $\cY^{u}$ is a derived quantity in this framework, cf.\,\eqref{cY:inversion}. In order to consider the seesaw regime, we filtered the points for which the singular values of 
$\sqrt{\cM^{Tu}\cM^{Tu\,\dagger}}$ are at least one order of magnitude larger than those of $\sqrt{\cM^T\cM^{T\,\dagger}}$. Moreover, we also applied a filter for perturbativity, allowing only points for which $\hat{\cY}^u_{ii}<4\pi$. We should notice that, for 2 NB-VLQs, Eq.\,(\ref{n=3:epsilon}) is not defined, while $\epsilon^{-1}\equiv{\cM^{T}}^{-1}{\cM^{Tu}}$ is. Thus, to make contact with the plots of the previous section, we will define $\lambda^{\epsilon}$ as the inverse of the singular values of $\epsilon^{-1}$.
\begin{figure}[ht!]
    \centering
    \includegraphics[scale=0.7]{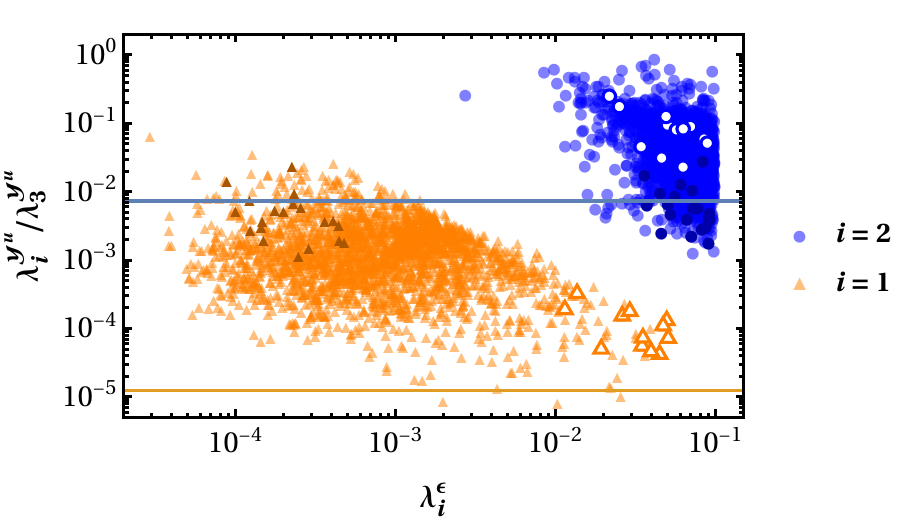}
\caption{
Singular values $\lambda_i^{\cY^u}$ as a function of singular values $\lambda_i^{\eps}$.
The hollow points comply with $\lambda_{1}^{\epsilon}/\lambda_{2}^{\epsilon}>0.5$, while for the darker points we require $\lambda_{1}^{\cY^u}/\lambda_{2}^{\cY^u}>0.6$. The orange (blue) line corresponds to the ratio between the up (charm) Yukawa to the top Yukawa, in the SM.}
\label{fig:2vlq_ratio}
\end{figure}

In figure~\ref{fig:2vlq_ratio}, we show the ratio among the two smallest singular values of $\cY^{u}$ to the third, against $\lambda^{\epsilon}_i$. As the ratio $\lambda_{1}^{y^{u}}/\lambda_{3}^{y^{u}}$ approaches the SM value (orange line), the singular values of $\epsilon$ are at most one order of magnitude apart. In this regime, the observed hierarchy among the first and second family masses comes from the hierarchical structure of $\cY^{u}$. 
This feature can be more clearly seen when considering the hollow points, which comply with $\lambda_{1}^{\epsilon}/\lambda_{2}^{\epsilon}>0.5$.
On the other hand, as the singular values of $\epsilon$ differ by roughly 3 orders of magnitude (which is the observed hierarchy among the first and second up-quark family in the SM), the ratios $\lambda_{1}^{y^{u}}/\lambda_{3}^{y^{u}}$, $\lambda_{2}^{y^{u}}/\lambda_{3}^{y^{u}}$ approach similar values. More precisely, they are close to the blue line, which represents the ratio $m_{c}/m_{t}$. In this case, the observed hierarchy among up and charm quark masses resides in $\epsilon$. 
This feature can be seen for the darker points, selected with $\lambda_{1}^{\cY^u}/\lambda_{2}^{\cY^u}>0.6$, where $\lambda_{1}^{\epsilon}$ and $\lambda_{2}^{\epsilon}$ are at least two orders of magnitude apart.

\begin{figure}[h!]
\centering
\includegraphics[scale=0.7]{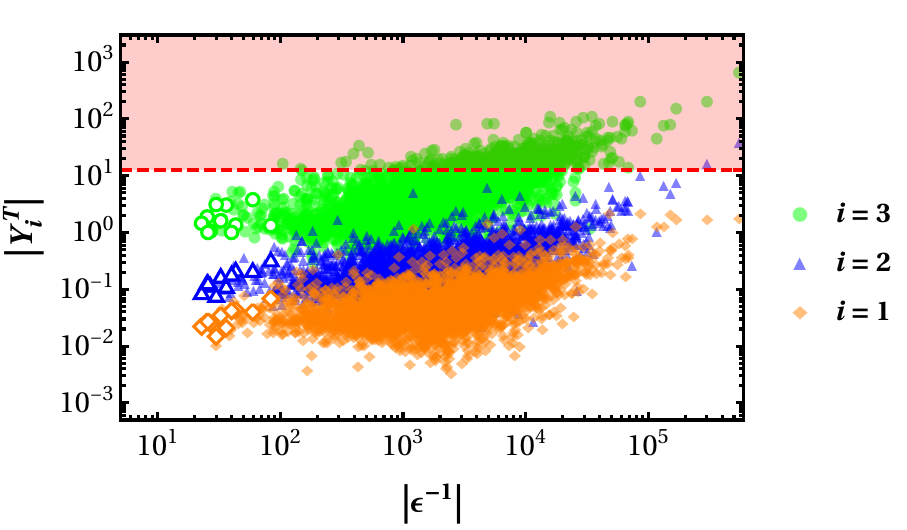}
\caption{
The norms $|Y^T_i|$ of VLQ Yukawa couplings in Eq.\,\eqref{def:YBi} as a function of the norm of $\eps^{-1}$ in Eq.\,\eqref{norm:eps-1}.
The red-shaded region denotes the perturbativity exclusion $|Y^T_i|<4\pi$.
}
\label{fig:YT-eps}
\end{figure}
\begin{figure}[h!]
\centering
\includegraphics[scale=0.7]{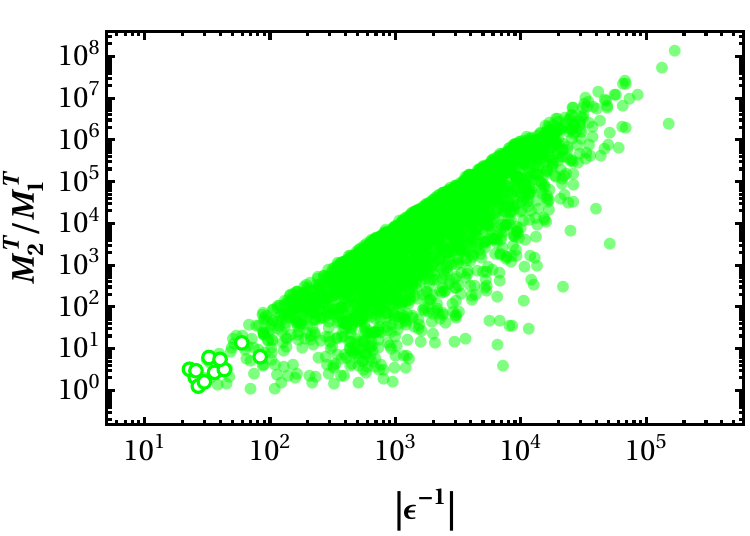}
\caption{
The ratio $M_2^T/M_1^T$ of the heavy quarks as a function of the norm of $\eps^{-1}$ in Eq.\,\eqref{norm:eps-1}.
}
\label{fig:MT-eps}
\end{figure}

In figure~\ref{fig:YT-eps}, we show how the norm $|Y^T_i|$ ($i=1,2,3$) varies in terms of the norm of $\epsilon^{-1}$, while in figure~\ref{fig:MT-eps} we illustrate the dependence of the ratio between the two VLQ masses in the same quantity. 
For comparison against the plots of the previous section, we added points that violate perturbativity in these figures. We notice that, for $|\eps^{-1}|>2\times10^{4}$, all green points are excluded setting an upper bound on this quantity.
Since the hollow points comply with  $\lambda_{1}^{\epsilon}/\lambda_{2}^{\epsilon}>0.5$, we see that if the hierarchy between the first and second up-quarks family resides in $\cY^{u}$, the norm of $\epsilon^{-1}$ attains its lowest
values. Similar patterns occur in the case of 3 VLQs. 

\begin{figure}[h!]
\centering
\includegraphics[scale=0.7]{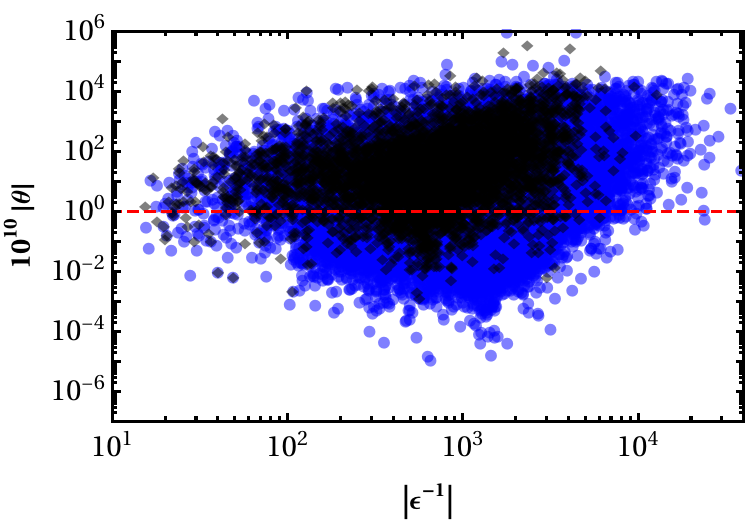}
\caption{
Estimate \eqref{invariants:n=2:up} of the irreducible 3-loop contributions to $\btheta$ against $|\eps^{-1}|$ in Eq.\,\eqref{norm:eps-1} for two up-type VLQs.
Points in black satisfy the seesaw limit condition while those in blue represent generic parameter choices.
}
\label{fig:theta-eps:2vlq}
\end{figure}

Finally, in Fig.\,\ref{fig:theta-eps:2vlq} we show the estimate to the irreducible contribution to $\btheta$ at 3-loops coming from the exchange of the VLQs against $|\eps^{-1}|$.
The following flavor invariant is expected to dominate for two or more up-type VLQs\,\cite{Valenti:2021xjp}:
\eqali{
\label{invariants:n=2:up}
\text{up-type $n_T\ge 2$: }\quad
\left(\frac{1}{16 \pi^2}\right)^3\im \text{Tr}\left( \left[\tY^{T^{\dag}} Y^u{Y^{u}}^\dagger \tilde{Y}^T, \tY^{T^{\dag}}\tilde{Y}^T\right]F\left({M^T}^\dag M^T\right)\right),}
where we are adopting the basis where $Y^T=\tilde{Y}^T$ in the basis where $Y^u=\hY^u$ is diagonal, and the function $F$ is defined below
\eqref{invariants:n=2:down}.
This estimation is expected to be corrected by order one prefactors depending on the model.
We compare generic points in blue, already quantified in Ref.\,\cite{Alves:2023cmw}, with the points in the seesaw limit where the singular values of $\cM^{Tu}$ is larger than the singular values of $\cM^T$ by at least one order of magnitude. 
We can see that plenty of points still obey $\btheta\lesssim 10^{-10}$ even if with a possible enhancement of an order of magnitude. 
As for the 3 VLQs case, we have filtered out the points that 
violate perturbativity.

\section{Conclusions}\label{sec:Con}

Although the SM is very successful in explaining a plethora of phenomena, there are still some puzzling issues, such as the strong CP problem and the observed fermion mass hierarchies, to cite a few. In this work, seeking a common solution to these problems, we have identified and investigated the ``seesaw limit'' of the Nelson-Barr mechanism. We have studied in detail how the introduction of three heavy VLQs, that mix with the Standard Model fields, can provide a natural mechanism to suppress the SM down quark masses in relation to the EW scale. In particular, we have seen that the source of CP violation in the quark sector, i.e., the SM-VLQ-mixing term $\cM^{Bd}$ in Eqs.\,(\ref{eq:Yukq}) and (\ref{eq:MdB}), is also behind the seesaw suppression of light quark masses. Since these VLQs are connected to the Nelson-Barr mechanism, not only the strong CP problem can be addressed, but also the couplings between them and the standard quarks are typically hierarchical. This in turn allows our model to naturally evade strong bounds from flavour observables coming from first and second quark families. We have also shown in a similar model with two up-type VLQs that the hierarchy among the SM up quarks can be partly explained by the hierarchy of the small seesaw parameters. For both cases, the irreducible 3-loop contributions to $\btheta$ severely restrict the parameter space of the model, especially for the case of three down-type VLQs. For the latter, such a restriction is weaker for the cases where the CP conserving Yukawa couplings $\cY^d$ of the down quarks in the CP basis are mostly responsible for the mass hierarchies.

%%%%%%%%%%%%%%%%%%%%%%%%%%%%%%%%%%%%%%%%%%%%%%%%%
\acknowledgments

This research was partially supported by the Conselho Nacional de Desenvolvimento Cient\'{\i}fico e Tecnol\'ogico (CNPq), by grants 308280/2023-7 (A.G.D.),
312866/2022-4 (C.C.N.), and 166523\slash2020-8 (A.L.C.).
Financial support by Funda\c{c}\~{a}o de Amparo \`{a} Pesquisa do Estado de S\~ao Paulo (FAPESP) is also acknowledged under the grant 2014/19164-6 (C.C.N.). J.L. is supported by the Spanish grants PID2020-113775GB-I00~(AEI/10.13039/501100011033) and Prometeo CIPROM/2021/054 (Generalitat Valenciana). A.L.C is supported by a postdoctoral fellowship from the Postdoctoral Researcher Program - Resolution GR/Unicamp No. 33/2023.

\bibliographystyle{apsrev4-1}

\bibliography{ref}
\end{document}